\documentclass[preprint,12pt]{elsarticle}

\usepackage{amssymb}

\usepackage{dcolumn}

\usepackage{graphicx}

\biboptions{sort&compress}

\journal{Annals of Physics}

\begin{document}

\begin{frontmatter}

\title{Non-Hermitian oscillators with $T_{d}$ symmetry}

\author{Paolo Amore\dag} \ead{paolo.amore@gmail.com}

\author{Francisco M.
Fern\'andez\ddag}\ead{fernande@quimica.unlp.edu.ar}

\author{Javier Garcia\ddag}

\address{\dag\ Facultad de Ciencias, CUICBAS,
Universidad de Colima, Bernal D\'{\i}az del Castillo 340, Colima,
Colima, Mexico}

\address{\ddag\ INIFTA (UNLP, CCT La Plata-CONICET), Divisi\'{o}n Qu\'{i}mica Te\'{o}rica,
Diag. 113 y 64 (S/N), Sucursal 4, Casilla de Correo 16, 1900 La
Plata, Argentina}

\begin{abstract}
We analyse some PT-symmetric oscillators with $T_{d}$ symmetry that depend
on a potential parameter $g$. We calculate the eigenvalues and
eigenfunctions for each irreducible representation and for a range of values
of $g$. Pairs of eigenvalues coalesce at exceptional points $g_c$; their
magnitude roughly decreasing with the magnitude of the eigenvalues. It is
difficult to estimate whether there is a phase transition at a nonzero value
of $g$ as conjectured in earlier papers. Group theory and perturbation
theory enable one to predict whether a given space-time symmetry leads to
real eigenvalues for sufficiently small nonzero values of $g$.
\end{abstract}


\end{frontmatter}

\section{Introduction}

\label{sec:intro}

In the last years there has been great interest in non-Hermitian
multidimensional oscillators with antiunitary symmetry $A=UK$,
where $U$ is a unitary operator and $K$ is the complex conjugation
operation. These Hamiltonians are of the form $H=H_{0}+igH^{\prime
}$, where $H_{0}$ is Hermitian, $UH_{0}U^{\dagger }=H_{0}$ and
$UH^{\prime }U^{\dagger }=-H^{\prime
}$\cite{BDMS01,NA02,N02,N05,BTZ06,KC08,W09,CIN10,BW12,HV13}. The
interest in these oscillators stems from the fact that they appear
to exhibit real eigenvalues for sufficiently small values of
$|g|$. As $g$ increases from $g=0$ two eigenvalues $E_{m}$ and
$E_{n}$ approach each other, coalesce at an exceptional point
$g_{c}$\cite{HS90,H00,HH01,H04} and become a pair of complex
conjugate numbers for $g>g_{c}$. At the exceptional point the
corresponding eigenvectors $\psi _{m}$ and $\psi _{n}$ are no
longer linearly independent\cite{HS90,H00,HH01,H04}. It is
commonly said that the system exhibits a PT-phase transition at
$g=g_{PT}>0$, where $g_{PT} $ is the exceptional point closest to
the origin\cite{BW12}. The eigenvalues $E_{m}$ of $H$ are real for
all $0\leq g<g_{PT}$, where the antiunitary symmetry remains
unbroken. Based on the multidimensional non-Hermitian oscillators
studied so far, Bender and Weir\cite{BW12} conjectured that the PT
phase transitions is a high-energy phenomenon.

Point-group symmetry (PGS)\cite{T64,C90} proved useful for the
study of a class of multidimensional anharmonic
oscillators\cite{PE81a,PE81b}. Klaiman and Cederbaum\cite{KC08}
applied PGS to non-Hermitian Hamiltonians chosen so that the point
group $G$ for $H$ is a subgroup of the point group $G_{0}$ for
$H_{0}$. They restricted their study to Abelian groups, which
exhibit only one-dimensional irreducible representations (irreps),
and Hermitian operators $H_{0}$ with no degenerate states. All
such examples exhibit real eigenvalues for sufficiently small
values of $|g|$. One of their goals was to predict the symmetry of
the eigenfunctions associated to the eigenvalues that coalesce at
the exceptional points and coined the term space-time (ST)
symmetry that refers to a class of antiunitary symmetries that
contain the PT symmetry as a particular case. Strictly speaking we
refer to PT symmetry when $U=P$,
$P:(\mathbf{x},\mathbf{p})\rightarrow (-\mathbf{x},-\mathbf{p})$,
where $\mathbf{x}$ and $\mathbf{p}$ are the collections of
coordinate and momenta operators, respectively.

The main interest in the studies of PT-symmetric multidimensional
oscillators just mentioned has been to enlarge the class of
non-Hermitian Hamiltonians that exhibit real spectra, at least for
some values of the potential parameter $g$. On the other hand, by
means of PGS Fern\'{a}ndez and Garcia\cite{FG14,FG14b} found some
examples of ST-symmetric multidimensional models that exhibit
complex eigenvalues for $g>0$ so that the phase transition takes
place at the trivial Hermitian limit $g_{PT}=0$. Their results
suggest that the more general ST symmetry is not as robust as the
PT one and contradict some of the conjectures put forward by
Klaiman and Cederbaum\cite{KC08} based on PGS. By means of PGS and
perturbation theory we have considerably improved the results,
arguments and conclusions of those earlier papers and also found a
greater class of ST-symmetric multidimensional models with broken
ST symmetry for all values of $g \neq 0$ \cite{AFG14}. Those
results show in a more clearly way that the conjecture of Klaiman
and Cederbaum does not apply to the general case where the
Hermitian Hamiltonian $H_{0}$ may exhibit degenerate states.

The purpose of this paper is the study of some tri-dimensional
non-Hermitian oscillators by means of PGS. In
section~\ref{sec:diagonalization} we discuss the diagonalization
of the matrix representation of the Hamiltonian operator in
symmetry-adapted basis sets. By means of PGS and perturbation
theory we develop a straightforward strategy that appears to be
suitable for determining whether the Hamiltonian will have real
eigenvalues for sufficiently small nonzero values of the parameter
$g$. In section~\ref {sec:example1} we choose a non-Hermitian
oscillator discussed earlier by Bender and Weir\cite{BW12} as an
illustrative example and exploit the fact that it exhibits $T_{d}$
symmetry. In section~\ref{sec:example2} we discuss a non-Hermitian
oscillator where $H_{0}$ and $H^{\prime }$ exhibits symmetry
$O_{h}$ and $T_{d}$, respectively. Finally, in
section~\ref{sec:conclusions} we draw conclusions.

\section{Diagonalization}

\label{sec:diagonalization}

Several approaches have been applied to the calculation of the
spectra of the ST-symmetric multidimensional oscillators: the
diagonalization method\cite{BDMS01,NA02,N02,N05,W09,BW12},
perturbation theory\cite {BDMS01,N02,N05,W09}, classical and
semiclassical approaches\cite {BDMS01,NA02}, among
others\cite{W09,HV13}. The diagonalization method consists of
expanding the eigenfunctions $\psi $ of $H$ as linear combinations
of a suitable basis set $B=\{f_{1},f_{2},\ldots \}$ %
\begin{equation}
\psi =\sum_{j}c_{j}f_{j}
\end{equation}
and then diagonalizing an $N\times N$ matrix representation of the
Hamiltonian $\mathbf{H}$ with elements $\left\langle f_{i}\right| H\left|
f_{j}\right\rangle $, where $\left\langle f\right| \left. g\right\rangle $
stands for the c-product\cite{M11}. Such matrices are complex and symmetric $%
\left\langle f_{i}\right| H\left| f_{j}\right\rangle =\left\langle
f_{j}\right| H\left| f_{i}\right\rangle $ but obviously not Hermitian.

In this paper we take into account that the non-Hermitian multidimensional
oscillators exhibit PGS and choose basis sets adapted to the irreps of the
point group $G$ of $H$. In this way we can split the matrix representation $%
\mathbf{H}$ into representations $\mathbf{H}^{S}$ for each symmetry $S$. The
eigenfunctions of $H$ are bases for the irreps of $G$ and can be written as
linear combinations
\begin{equation}
\psi ^{S}=\sum_{j}c_{j}^{S}f_{j}^{S}  \label{eq:psi^S}
\end{equation}
of the elements of the symmetry-adapted basis sets $B^{S}=%
\{f_{1}^{S},f_{2}^{S},\ldots \}$. The matrix elements of $\mathbf{H}^{S}$
are given by $\left\langle f_{i}^{S}\right| H\left| f_{j}^{S}\right\rangle $
and the separate treatment of each symmetry is justified by the fact that $%
\left\langle f_{i}^{S}\right| H\left| f_{j}^{S^{\prime }}\right\rangle =0$
if $S\neq S^{\prime }$\cite{T64,C90}. That is to say: functions of different
symmetry do not mix.

The construction of symmetry-adapted basis sets is straightforward and is
described in most textbooks on group theory\cite{T64,C90}. One applies a
projection operator $P^{S}$ to a basis function $f_{j}$ and obtains a
symmetry-adapted function $u_{j}^{S}$. If the irrep $S$ is one-dimensional
it is only necessary to normalize the resulting function $u_{j}^{S}$;
otherwise it may be necessary to combine two or more functions $u_{j}^{S}$
to obtain a set of orthonormal functions\cite{T64,C90}.

In what follows we apply this approach to two non-Hermitian
three-dimensional oscillators of the form
\begin{equation}
H=H_{0}+igH^{\prime }  \label{eq:H_gen}
\end{equation}
where $H_{0}$ is Hermitian and $g$ is real. In particular, we consider the
case that both $H_{0}$ and $H$ exhibit eigenspaces of dimension greater than
one.

In the examples discussed in this paper the symmetry of $H_{0}$ is given by
the point group $G_{0}=\left\{ U_{1},U_{2},\ldots ,U_{m}\right\} $: $%
U_{i}H_{0}U_{i}^{-1}=H_{0}$. If $H^{\prime }$ is invariant under the
operations of a subgroup $G=\left\{ W_{1},W_{2},\ldots ,W_{k}\right\} $ of $%
G_{0}$ ($W_{i}H^{\prime }W_{i}^{-1}=H^{\prime }$) then $H$ is invariant
under the operations of the point group $G$. Suppose that there exists a
unitary operator $U_{a}\in $ $G_{0}\backslash G$ with the following
properties: i) it forms a class by itself (that is to say: $%
U_{i}U_{a}U_{i}^{-1}=U_{a}$, $i=1,2,\ldots ,m$) so that $U_{a}^{-1}=U_{a}$,
ii) it changes the sign of $H^{\prime }$ $U_{a}H^{\prime
}U_{a}^{-1}=-H^{\prime }$. Under these conditions $H$ exhibits the
antiunitary symmetry given by $A=U_{a}K$, $AHA^{-1}=H$, where $K$ is the
complex conjugation operation introduced earlier.

If $\psi _{m}^{(0)}$ is an eigenfunction of $H_0$ with eigenvalue $%
E_{m}^{(0)}$ then $U_{a}\psi _{m}^{(0)}=\sigma _{m}\psi _{m}^{(0)}$, where $%
\sigma _{m}=\pm 1$, as follows from $[H,U_{a}]=0$ and $U_{a}^{2}=1$.
Therefore,
\begin{equation}
\left\langle \psi _{m}^{(0)}\right| H^{\prime }\left| \psi
_{n}^{(0)}\right\rangle =0,
\end{equation}
if $\sigma _{m}\sigma _{n}=1$.

It was shown in our earlier papers that complex eigenvalues appear for
sufficiently small values of $|g|$ when $H_{0}$ exhibits degenerate
eigenfunctions and at least one of the perturbation corrections of first
order produced by $H^{\prime }$ is nonzero\cite{FG14,FG14b,AFG14}. The
degenerate eigenfunctions of $H_{0}$%
\begin{equation}
H_{0}\psi _{m,k}^{(0)}=E_{m}^{(0)}\psi _{m,k}^{(0)},\;k=1,2,\ldots ,\nu _{m},
\end{equation}
exhibit the same behaviour with respect to $U_{a}$: $U_{a}\psi
_{m,k}^{(0)}=\sigma _{m}\psi _{m,k}^{(0)}$, so that
\begin{equation}
\left\langle \psi _{m,k}^{(0)}\right| H^{\prime }\left| \psi
_{n,l}^{(0)}\right\rangle =0\;k,l=1,2,\ldots ,\nu _{m},
\end{equation}
and all the perturbation corrections of first order vanish (see reference%
\cite{AFG14} for more details).

The main conclusion drawn from the discussion above is that the space-time
symmetry given by $A$ may not be broken when the space transformation given
by $U_{a}\in G_{0}$ forms a class by itself. This conjecture is confirmed by
all the examples discussed in our earlier paper\cite{AFG14} where we
concluded that the inversion operation $\hat{\imath}:(x,y,z)\rightarrow
(-x,-y,-z)$ is a suitable choice for $U_{a}$. Note that in all the point
groups $\hat{\imath}$ forms a class by itself\cite{T64,C90}.

\section{Example 1}

\label{sec:example1}

As a first example we choose the non-Hermitian oscillator
\begin{equation}
H=p_{x}^{2}+p_{y}^{2}+p_{z}^{2}+x^{2}+y^{2}+z^{2}+igxyz  \label{eq:H_BW}
\end{equation}
studied by Bender and Weir\cite{BW12}. When $g=0$ the resulting isotropic
harmonic oscillator $H_{0}$ may be described by the 3D rotation group (the
group of all rotations about the origin of the three-dimensional Euclidean
space $R^{3}$ under the operation of composition). Its eigenfuctions in
Cartesian coordinates are
\begin{equation}
\varphi _{m,n,k}(x,y,z)=\phi _{m}(x)\phi _{n}(y)\phi
_{k}(z),\;m,n,k=0,1,\ldots ,  \label{eq:varphi_mnk}
\end{equation}
where $\phi _{j}(q)$ is an eigenfunction of $H_{HO}=p_{q}^{2}+q^{2}$, and
the corresponding eigenvalues
\begin{equation}
E_{mnk}^{(0)}=2\nu +3,\;\nu =m+n+k,
\end{equation}
are $(\nu +1)(\nu +2)/2$-fold degenerate. When $g\neq 0$ the symmetry of the
model is determined by $H^{\prime }=xyz$ and the suitable point group is $%
T_{d}$. The corresponding character table is shown in Table~\ref{tab:Td}. In
this case the obvious choice is $U_{a}=\hat{\imath}\in G_{0}$ that satisfies
all the conditions outlined in section~\ref{sec:diagonalization}.

The application of the projection procedure outlined in section~\ref
{sec:diagonalization} to the eigenfunctions of $H_{0}$ yields the following
symmetry-adapted basis set for $G=T_{d}$%
\begin{eqnarray}
A_{1} &:&\left\{
\begin{array}{l}
\varphi _{2m,2m,2m} \\
\frac{1}{\sqrt{3}}\left( \varphi _{2m,2m,2n}+\varphi _{2m,2n,2m}+\varphi
_{2n,2m,2m}\right) \\
\frac{1}{\sqrt{6}}\left( \varphi _{2m,2n,2k}+\varphi _{2k,2m,2n}+\varphi
_{2n,2k,2m}+\varphi _{2k,2n,2m}+\varphi _{2m,2k,2n}+\varphi
_{2n,2m,2k}\right) \\
\varphi _{2m+1,2m+1,2m+1} \\
\frac{1}{\sqrt{3}}\left( \varphi _{2m+1,2m+1,2n+1}+\varphi
_{2m+1,2n+1,2m+1}+\varphi _{2n+1,2m+1,2m+1}\right) \\
\frac{1}{\sqrt{6}}\left( \varphi _{2m+1,2n+1,2k+1}+\varphi
_{2k+1,2m+1,2n+1}+\varphi _{2n+1,2k+1,2m+1}+\varphi _{2k+1,2n+1,2m+1}\right.
\\
\left. +\varphi _{2m+1,2k+1,2n+1}+\varphi _{2n+1,2m+1,2k+1}\right)
\end{array}
\right.  \nonumber \\
A_{2} &:&\left\{
\begin{array}{l}
\frac{1}{\sqrt{6}}\left( \varphi _{2m,2n,2k}+\varphi _{2k,2m,2n}+\varphi
_{2n,2k,2m}-\varphi _{2k,2n,2m}-\varphi _{2m,2k,2n}-\varphi
_{2n,2m,2k}\right) \\
\frac{1}{\sqrt{6}}\left( \varphi _{2m+1,2n+1,2k+1}+\varphi
_{2k+1,2m+1,2n+1}+\varphi _{2n+1,2k+1,2m+1}-\varphi _{2k+1,2n+1,2m+1}\right.
\\
\left. -\varphi _{2m+1,2k+1,2n+1}-\varphi _{2n+1,2m+1,2k+1}\right)
\end{array}
\right.  \nonumber \\
&&
\end{eqnarray}
\begin{equation}
E:\left\{
\begin{array}{l}
\left\{ \frac{1}{\sqrt{6}}\left( 2\varphi _{2n,2m,2m}-\varphi
_{2m,2n,2m}-\varphi _{2m,2m,2n}\right) ,\frac{1}{\sqrt{2}}\left( \varphi
_{2m,2n,2m}-\varphi _{2m,2m,2n}\right) \right\} \\
\left\{ \frac{1}{\sqrt{6}}\left( 2\varphi _{2m,2n,2k}-\varphi
_{2k,2m,2n}-\varphi _{2n,2k,2m}\right) ,\frac{1}{\sqrt{2}}\left( \varphi
_{2k,2m,2n}-\varphi _{2n,2k,2m}\right) \right\} \\
\left\{ \frac{1}{\sqrt{6}}\left( 2\varphi _{2n,2m,2k}-\varphi
_{2k,2n,2m}-\varphi _{2m,2k,2n}\right) ,\frac{1}{\sqrt{2}}\left( \varphi
_{2k,2n,2m}-\varphi _{2m,2k,2n}\right) \right\} \\
\left\{
\begin{array}{c}
\frac{1}{\sqrt{6}}\left( 2\varphi _{2n+1,2m+1,2m+1}-\varphi
_{2m+1,2n+1,2m+1}-\varphi _{2m+1,2m+1,2n+1}\right) , \\
\frac{1}{\sqrt{2}}\left( \varphi _{2m+1,2n+1,2m+1}-\varphi
_{2m+1,2m+1,2n+1}\right)
\end{array}
\right\} \\
\left\{
\begin{array}{c}
\frac{1}{\sqrt{6}}\left( 2\varphi _{2m+1,2n+1,2k+1}-\varphi
_{2k+1,2m+1,2n+1}-\varphi _{2n+1,2k+1,2m+1}\right) , \\
\frac{1}{\sqrt{2}}\left( \varphi _{2k+1,2m+1,2n+1}-\varphi
_{2n+1,2k+1,2m+1}\right)
\end{array}
\right\} \\
\left\{
\begin{array}{c}
\frac{1}{\sqrt{6}}\left( 2\varphi _{2n+1,2m+1,2k+1}-\varphi
_{2k+1,2n+1,2m+1}-\varphi _{2m+1,2k+1,2n+1}\right) , \\
\frac{1}{\sqrt{2}}\left( \varphi _{2k+1,2n+1,2m+1}-\varphi
_{2m+1,2k+1,2n+1}\right)
\end{array}
\right\}
\end{array}
\right.
\end{equation}
\begin{equation}
T_{1}:\left\{
\begin{array}{l}
\left\{
\begin{array}{c}
\frac{1}{\sqrt{2}}\left( \varphi _{2m+1,2n,2k+1}-\varphi
_{2k+1,2n,2m+1}\right) ,\frac{1}{\sqrt{2}}\left( \varphi
_{2k+1,2m+1,2n}-\varphi _{2m+1,2k+1,2n}\right) , \\
\frac{1}{\sqrt{2}}\left( \varphi _{2n,2k+1,2m+1}-\varphi
_{2n,2m+1,2k+1}\right)
\end{array}
\right\} \\
\left\{
\begin{array}{c}
\frac{1}{\sqrt{2}}\left( \varphi _{2m,2n+1,2k}-\varphi _{2k,2n+1,2m}\right) ,%
\frac{1}{\sqrt{2}}\left( \varphi _{2k,2m,2n+1}-\varphi _{2m,2k,2n+1}\right) ,
\\
\frac{1}{\sqrt{2}}\left( \varphi _{2n+1,2k,2m}-\varphi _{2n+1,2m,2k}\right)
\end{array}
\right\}
\end{array}
\right.
\end{equation}
\begin{equation}
T_{2}:\left\{
\begin{array}{c}
\begin{array}{c}
\begin{array}{c}
\left\{ \varphi _{2m+1,2n,2n},\varphi _{2n,2m+1,2n},\varphi
_{2n,2n,2m+1}\right\} \\
\left\{ \varphi _{2m,2n+1,2n+1},\varphi _{2n+1,2m,2n+1},\varphi
_{2n+1,2n+1,2m}\right\}
\end{array}
\\
\left\{
\begin{array}{c}
\frac{1}{\sqrt{2}}\left( \varphi _{2m+1,2n,2k+1}+\varphi
_{2k+1,2n,2m+1}\right) ,\frac{1}{\sqrt{2}}\left( \varphi
_{2k+1,2m+1,2n}+\varphi _{2m+1,2k+1,2n}\right) , \\
\frac{1}{\sqrt{2}}\left( \varphi _{2n,2k+1,2m+1}+\varphi
_{2n,2m+1,2k+1}\right)
\end{array}
\right\}
\end{array}
\\
\left\{
\begin{array}{c}
\frac{1}{\sqrt{2}}\left( \varphi _{2m,2n+1,2k}+\varphi _{2k,2n+1,2m}\right) ,%
\frac{1}{\sqrt{2}}\left( \varphi _{2k,2m,2n+1}+\varphi _{2m,2k,2n+1}\right) ,
\\
\frac{1}{\sqrt{2}}\left( \varphi _{2n+1,2k,2m}+\varphi _{2n+1,2m,2k}\right)
\end{array}
\right\}
\end{array}
\right.
\end{equation}
By means of projection operators one can also prove that the perturbation $%
H^{\prime }=xyz$ splits the degenerate states of the three-dimensional
Harmonic oscillator $H_{0}$ in the following way:
\begin{eqnarray}
\{2n,2n,2n\} &\rightarrow &A_{1}  \nonumber \\
\{2n+1,2m,2m\}_{P} &\rightarrow &T_{2}  \nonumber \\
\{2n+1,2n+1,2m\}_{P} &\rightarrow &T_{2}  \nonumber \\
\{2n,2m,2m\}_{P} &\rightarrow &A_{1},E  \nonumber \\
\{2n+1,2n+1,2n+1\} &\rightarrow &A_{1}  \nonumber \\
\{2n,2m,2k+1\}_{P} &\rightarrow &T_{1},T_{2}  \nonumber \\
\{2n,2m+1,2k+1\}_{P} &\rightarrow &T_{1},T_{2}  \nonumber \\
\{2n,2m,2k\}_{P} &\rightarrow &A_{1},A_{2},E,E  \nonumber \\
\{2n+1,2m+1,2m+1\}_{P} &\rightarrow &A_{1},E  \nonumber \\
\{2n+1,2m+1,2k+1\}_{P} &\rightarrow &A_{1},A_{2},E,E,  \label{eq:R3->Td}
\end{eqnarray}
where $\{i,j,k\}_{P}$ denotes all the distinct permutations of the labels $i$%
, $j$ and $k$.

Bender and Weir\cite{BW12} diagonalized truncated matrix representations $%
\mathbf{H}$ of the Hamiltonian operator of dimension $20^{3}\times 20^{3}$, $%
25^{3}\times 25^{3}$ and $30^{3}\times 30^{3}$ in order to estimate the
accuracy of their results. They resorted to well known efficient
diagonalization routines for sparse matrices. Here, we diagonalize matrix
representations $\mathbf{H}^{S}$ for $S=A_{1},A_{2},E,T_{1},T_{2}$. This
splitting reduces the dimension of the matrices required for a given
accuracy and also enables a clearer interpretation and discussion of the
results. In this paper we carried out all the calculations with matrices of
dimension $5000\times 5000$ for each irrep. Comparison of such results for $%
g=1$ with those coming from a calculation with matrices of dimension $%
10000\times 10000$ did not show any relevant difference for present purposes
and discussion.

\begin{figure}[]
\begin{center}
\includegraphics[width=6cm]{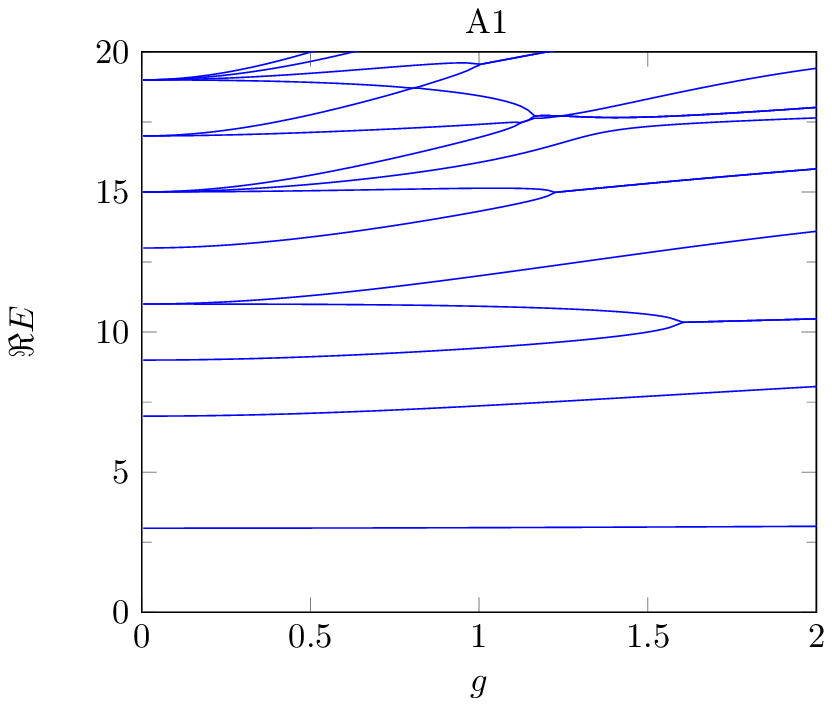} %
\includegraphics[width=6cm]{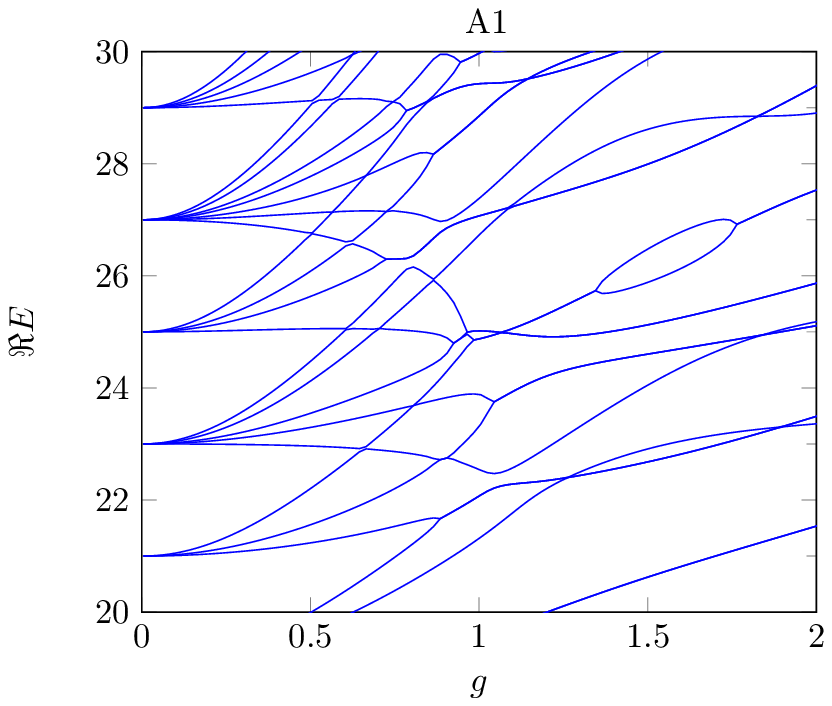} %
\includegraphics[width=6cm]{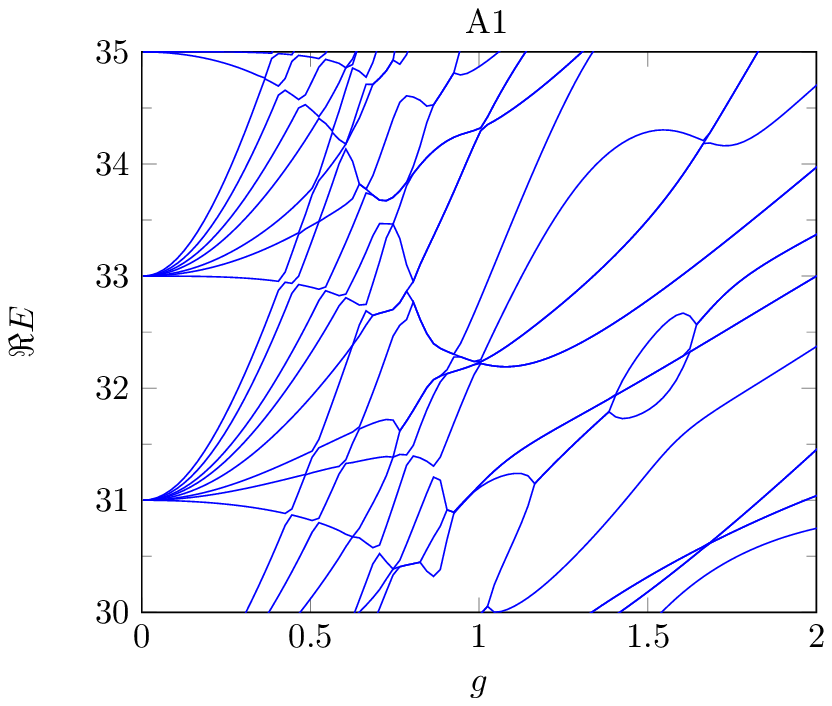}
\par
\end{center}
\caption{Real parts of the eigenvalues of symmetry $A_1$ of the Hamiltonian
operator $H=p_x^2+p_y^2+p_z^2+x^2+y^2+z^2+igxyz$}
\label{fig:HO_A1}
\end{figure}

\begin{figure}[]
\begin{center}
\includegraphics[width=6cm]{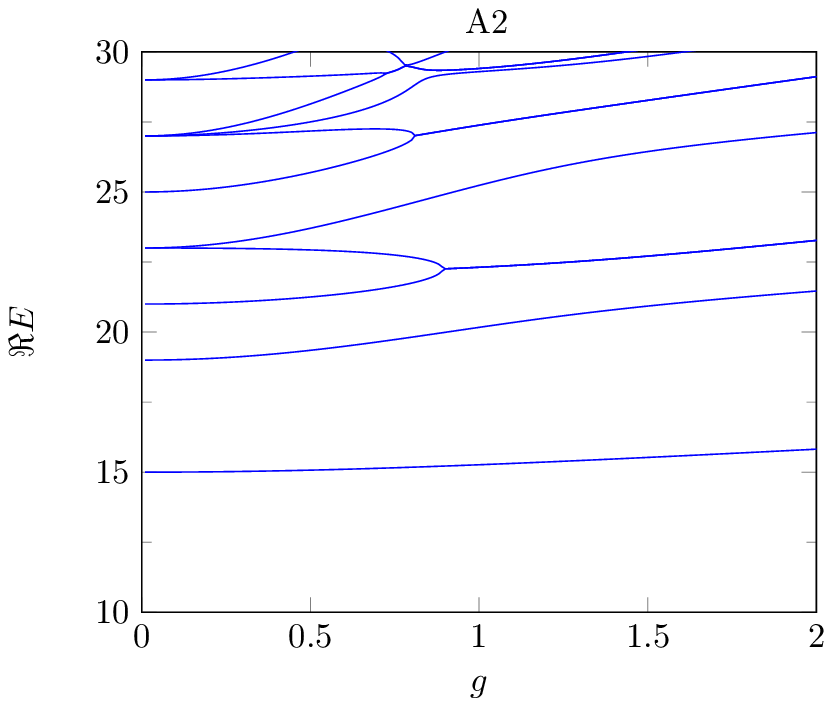} %
\includegraphics[width=6cm]{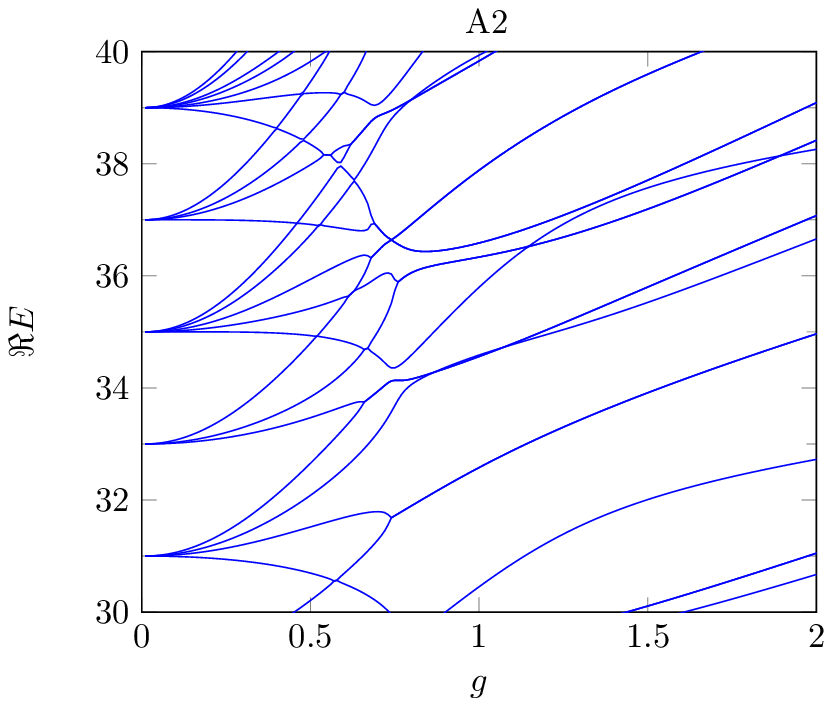} %
\includegraphics[width=6cm]{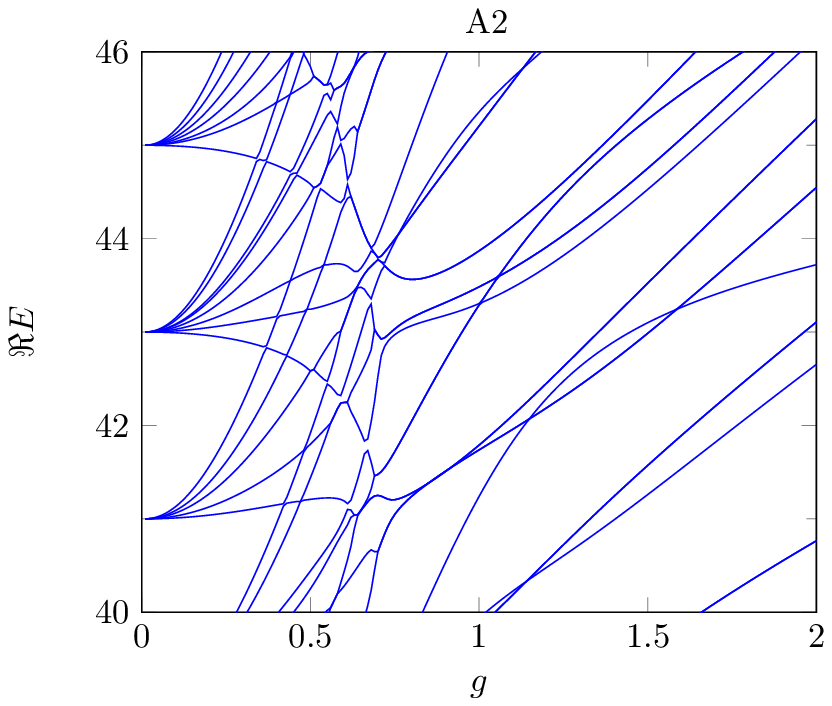}
\par
\end{center}
\caption{Real parts of the eigenvalues of symmetry $A_2$ the Hamiltonian
operator $H=p_x^2+p_y^2+p_z^2+x^2+y^2+z^2+igxyz$}
\label{fig:HO_A2}
\end{figure}

\begin{figure}[]
\begin{center}
\includegraphics[width=6cm]{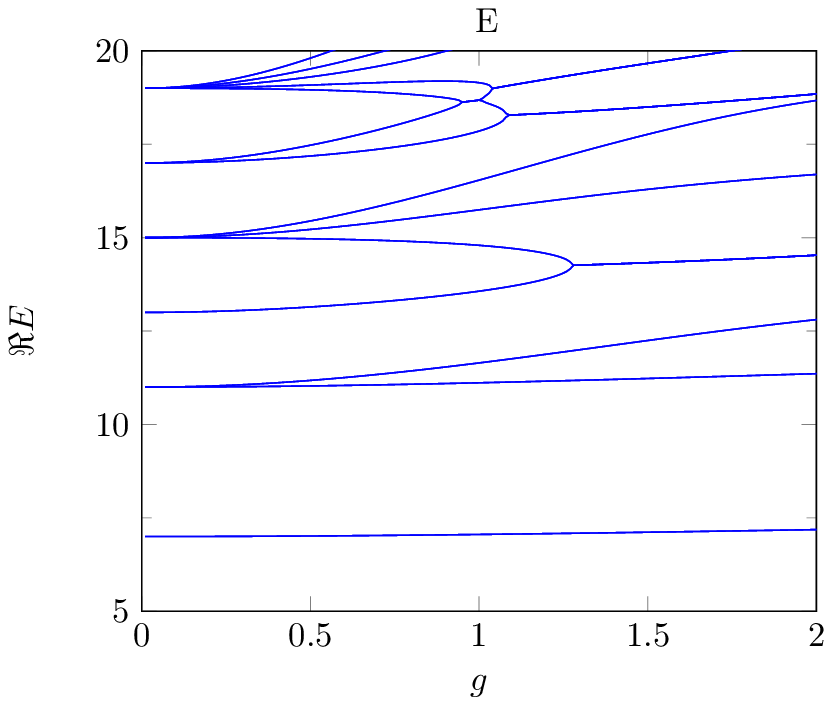} %
\includegraphics[width=6cm]{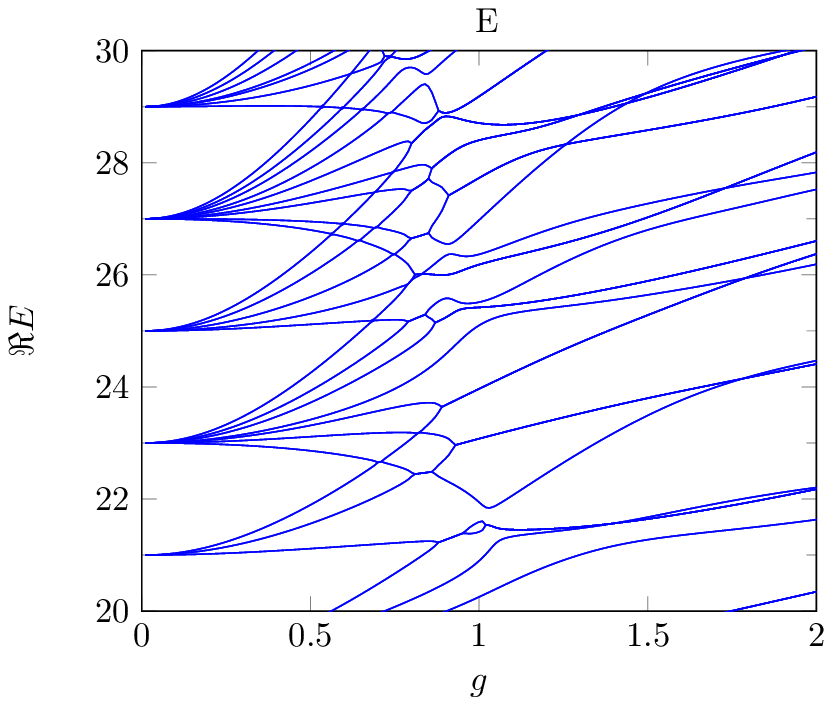}
\par
\end{center}
\caption{Real parts of the eigenvalues of symmetry $E$ the Hamiltonian
operator $H=p_x^2+p_y^2+p_z^2+x^2+y^2+z^2+igxyz$}
\label{fig:HO_E}
\end{figure}

\begin{figure}[]
\begin{center}
\includegraphics[width=6cm]{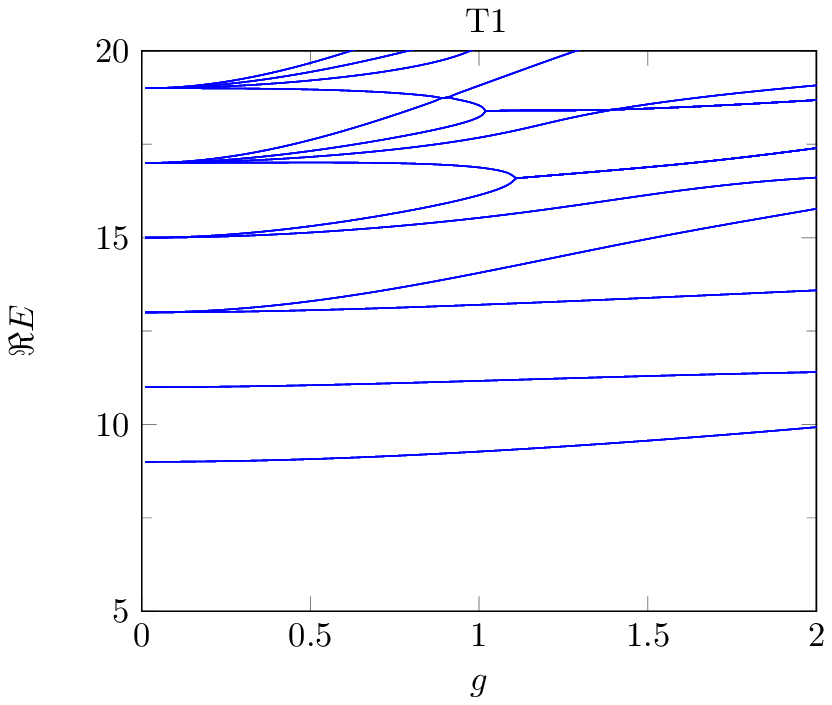} %
\includegraphics[width=6cm]{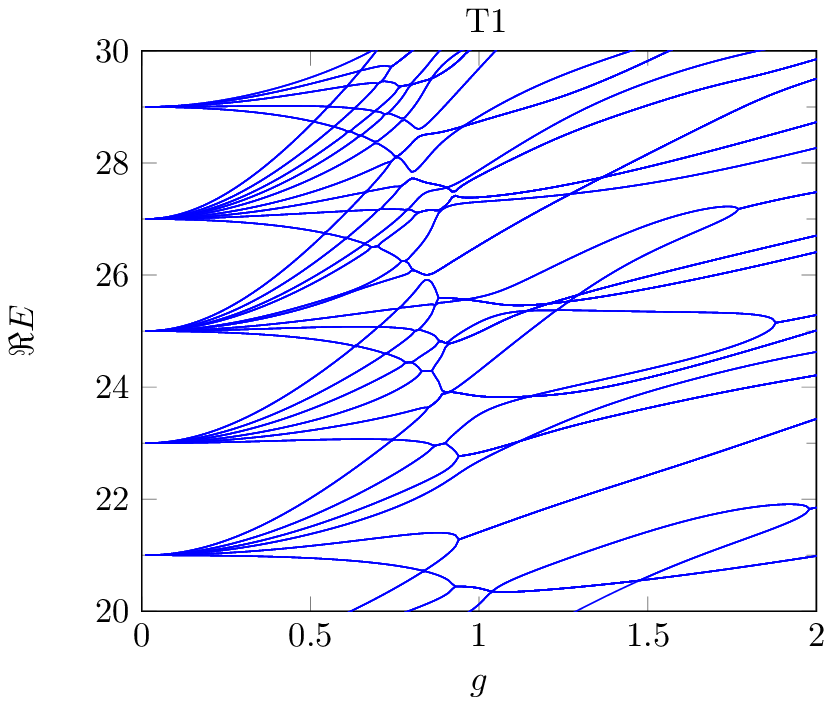}
\par
\end{center}
\caption{Real parts of the eigenvalues of symmetry $T_1$ the Hamiltonian
operator $H=p_x^2+p_y^2+p_z^2+x^2+y^2+z^2+igxyz$}
\label{fig:HO_T1}
\end{figure}

\begin{figure}[]
\begin{center}
\includegraphics[width=6cm]{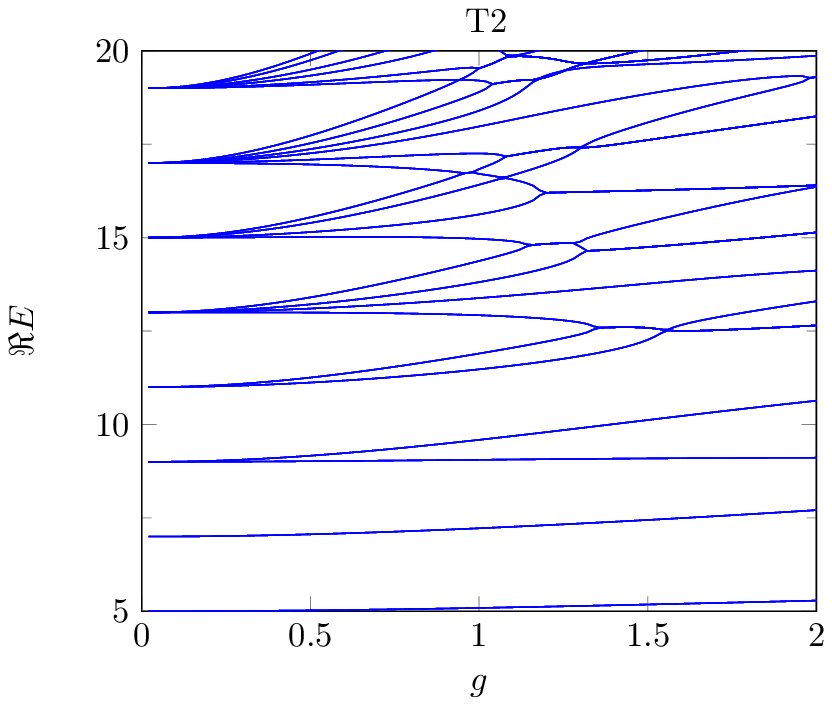} %
\includegraphics[width=6cm]{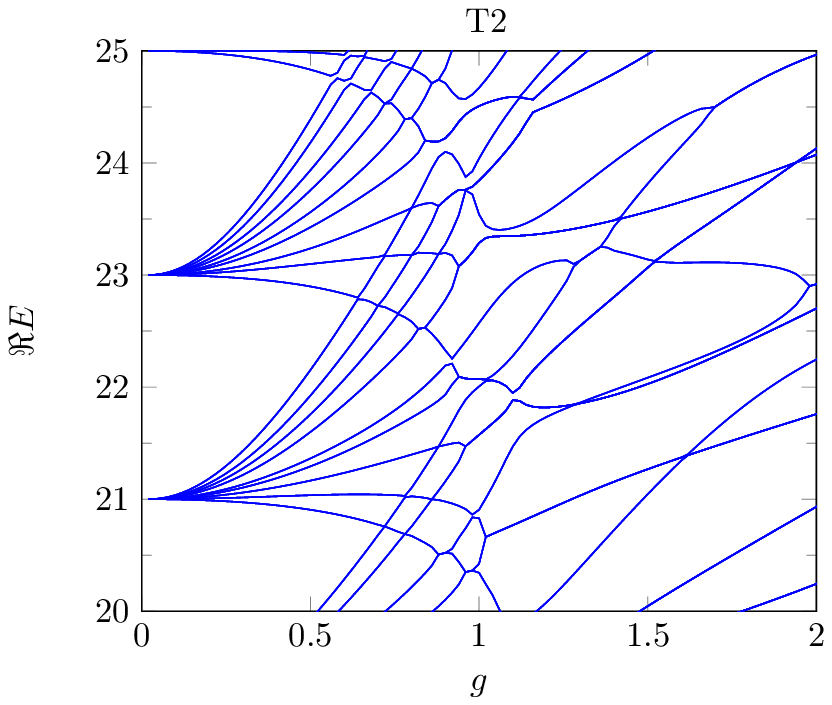}
\par
\end{center}
\caption{Real parts of the eigenvalues of symmetry $T_2$ the Hamiltonian
operator $H=p_x^2+p_y^2+p_z^2+x^2+y^2+z^2+igxyz$}
\label{fig:HO_T2}
\end{figure}

Figures \ref{fig:HO_A1}-\ref{fig:HO_T2} show $\Re E(g)$ for the five irreps.
For clarity we split every case into two or three energy intervals where we
can appreciate the occurrence of crossings, coalescence of eigenvalues at
exceptional points and even what appear to be avoided crossings. Because of
the scale used and the separation into irreps our figures reveal a reach
pattern of intertwined energy curves that one cannot easily discern when
plotting all the symmetries together\cite{BW12}.

\begin{figure}[]
\begin{center}
\includegraphics[width=6cm]{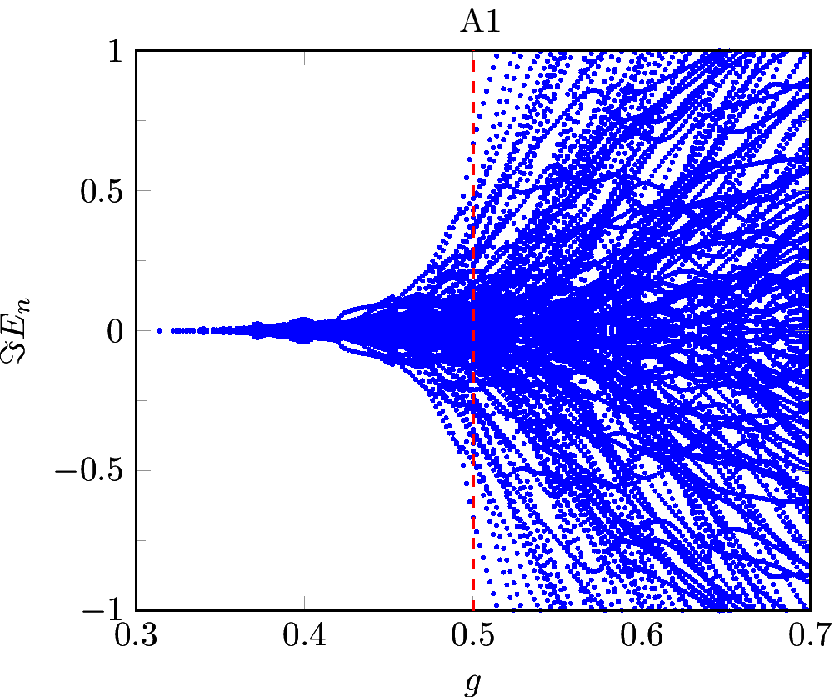} %
\includegraphics[width=6cm]{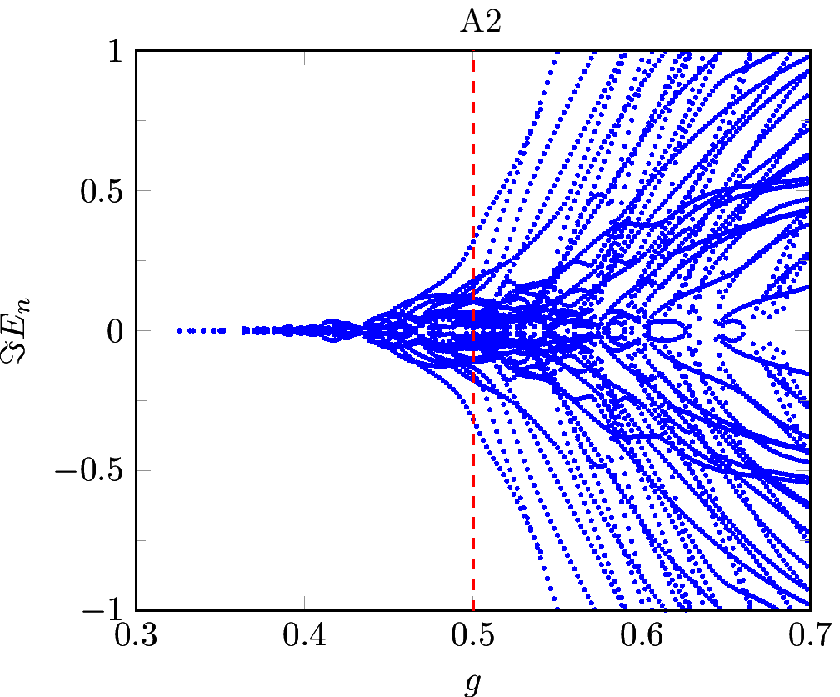} %
\includegraphics[width=6cm]{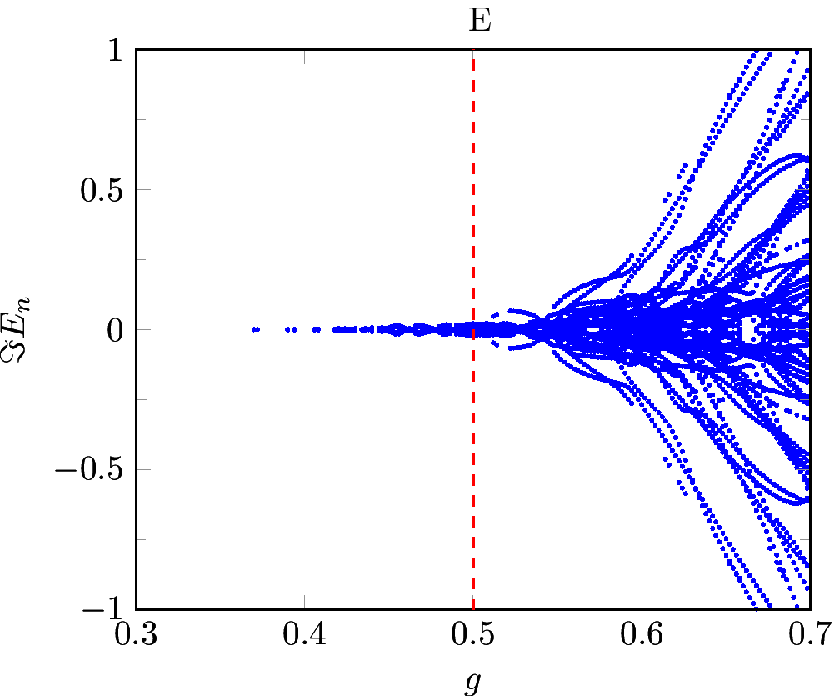} %
\includegraphics[width=6cm]{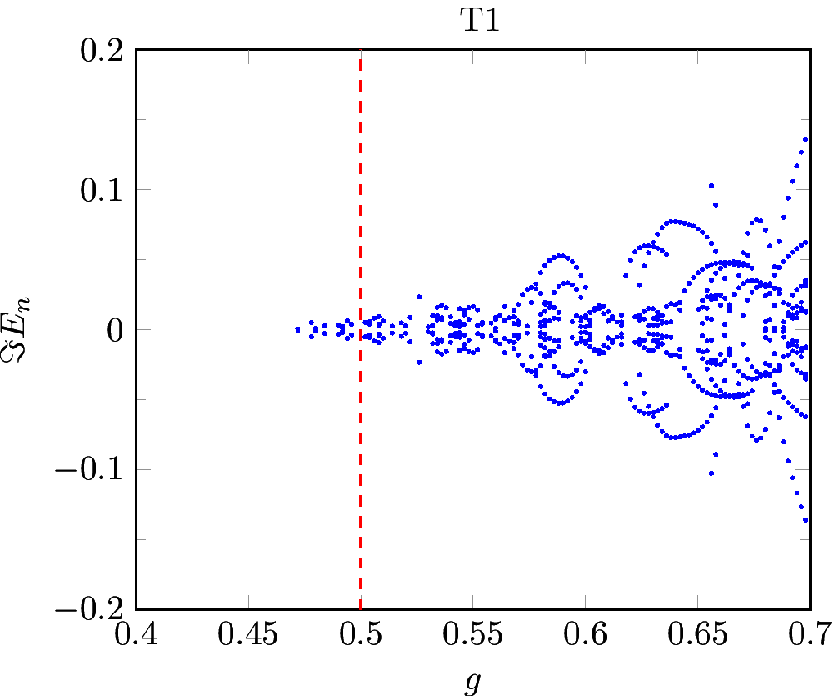} %
\includegraphics[width=6cm]{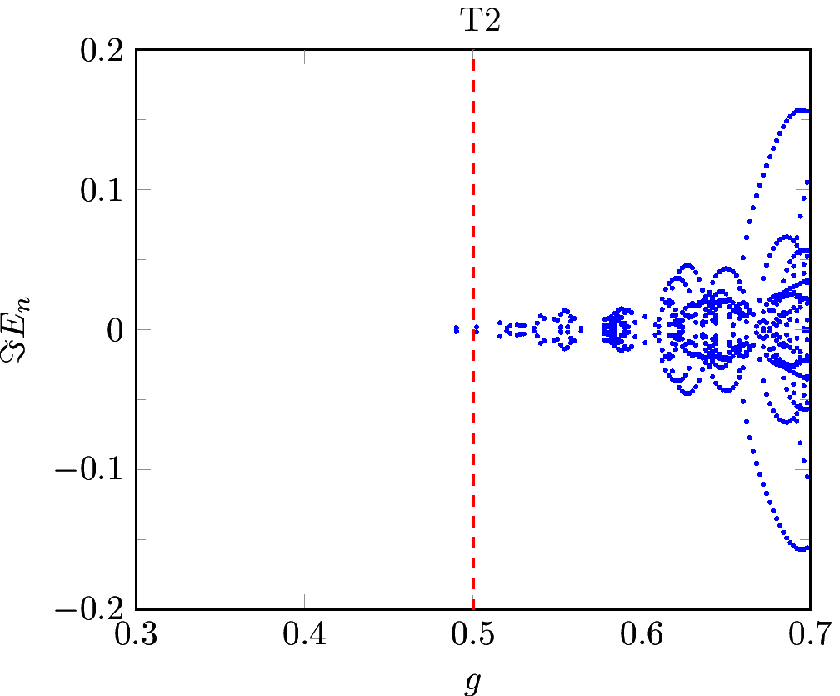}
\par
\end{center}
\caption{Imaginary parts of the eigenvalues of the Hamiltonian Operator $%
H=p_x^2+p_y^2+p_z^2+x^2+y^2+z^2+igxyz$}
\label{fig:HO_ImE}
\end{figure}

Bender and Weir\cite{BW12} estimated a phase transition near to $g\approx
0.25$ for their Hamiltonian $H^{BW}=H_{0}^{present}/2+igxyz$. The relation
between present exceptional points and those of Bender and Weir is therefore
$g_{c}^{present}=2g_{c}^{BW}$. Figure \ref{fig:HO_ImE} shows the imaginary
parts of the eigenvalues for each irrep. We appreciate that complex
eigenvalues appear for values of the parameter that are considerably smaller
than $g=0.5$; therefore, we cannot be sure that there is a phase transition
for this Hamiltonian. As the energy increases more exceptional points seems
to emerge closer to the origin.

All the figures in this paper have been produced by means of the Tikz package%
\cite{TikZ}.

\section{Example 2}

\label{sec:example2}

The non-Hermitian anharmonic oscillator
\begin{equation}
H=p_{x}^{2}+p_{y}^{2}+p_{z}^{2}+x^{4}+y^{4}+z^{4}+igxyz,  \label{eq:H_ours}
\end{equation}
is interesting because $H_{0}$ is invariant under the unitary operations of
the point group $O_{h}$ and $H$ is invariant under those of $T_{d}$.

If $\{i,j,k\}_{P}$ denotes all distinct permutations of the subscripts in
the eigenfunctions $\chi _{i\,j\,k}(x,y,z)=\rho _{i}(x)\rho _{j}(y)\rho
_{k}(z)$, $i,j,k=0,1,\ldots $, of $H_{0}$, then their symmetry is given by\
(see reference\cite{F13b,HCL13} for a discussion of an exactly solvable
quantum-mechanical problem with the same PGS):

\begin{equation}
\begin{array}{ll}
\{2n,2n,2n\} & A_{1g} \\
\{2n+1,2n+1,2n+1\} & A_{2u} \\
\{2n+1,2n+1,2m\}_{P} & T_{2g} \\
\{2n,2n,2m+1\}_{P} & T_{1u} \\
\{2n,2n,2m\}_{P} & A_{1g},E_{g} \\
\{2n+1,2n+1,2m+1\}_{P} & A_{2u},E_{u} \\
\{2n,2m,2k\}_{P} & A_{1g},A_{2g},E_{g},E_{g} \\
\{2n+1,2m+1,2k+1\}_{P} & A_{1u},A_{2u},E_{u},E_{u} \\
\{2n,2m,2k+1\}_{P} & T_{1u},T_{2u} \\
\{2n+1,2m+1,2k\}_{P} & T_{1g},T_{2g}
\end{array}
.  \label{eq:degeneracy_2}
\end{equation}
The character table for the point group $O_{h}$ is shown in Table~\ref
{tab:oh}. The dynamical symmetries that are responsible for the degeneracy
of eigenfunctions belonging to different irreps (which cannot be explained
by PGS) are given by the Hermitian operators
\begin{eqnarray}
O_{1} &=&2p_{x}^{2}+2x^{4}-p_{y}^{2}-y^{4}-p_{z}^{2}-z^{4}  \nonumber \\
O_{2} &=&2p_{y}^{2}+2y^{4}-p_{x}^{2}-x^{4}-p_{z}^{2}-z^{4}.
\end{eqnarray}
They belong to the irrep $E_{g}$ and commute with $H_0$. We easily obtain
them by straightforward application of the projection operator $P^{E_{g}}$
to the two pairs of functions $(x^{2},y^{2})$ and $(x^{4},y^{4})$ as
discussed elsewhere\cite{F13b}.

By means of projection operators we can prove that the eigenfunctions of $%
H_{0}$ transform into those of $H$ according to the following symmetry
scheme:
\begin{eqnarray}
A_{1g},\,A_{2u} &\rightarrow &A_{1}  \nonumber \\
A_{2g},\,A_{1u} &\rightarrow &A_{2}  \nonumber \\
E_{g},\,E_{u} &\rightarrow &E  \nonumber \\
T_{1g},\,T_{2u} &\rightarrow &T_{1}  \nonumber \\
T_{2g},\,T_{1u} &\rightarrow &T_{2}
\end{eqnarray}

\begin{figure}[]
\begin{center}
\includegraphics[width=6cm]{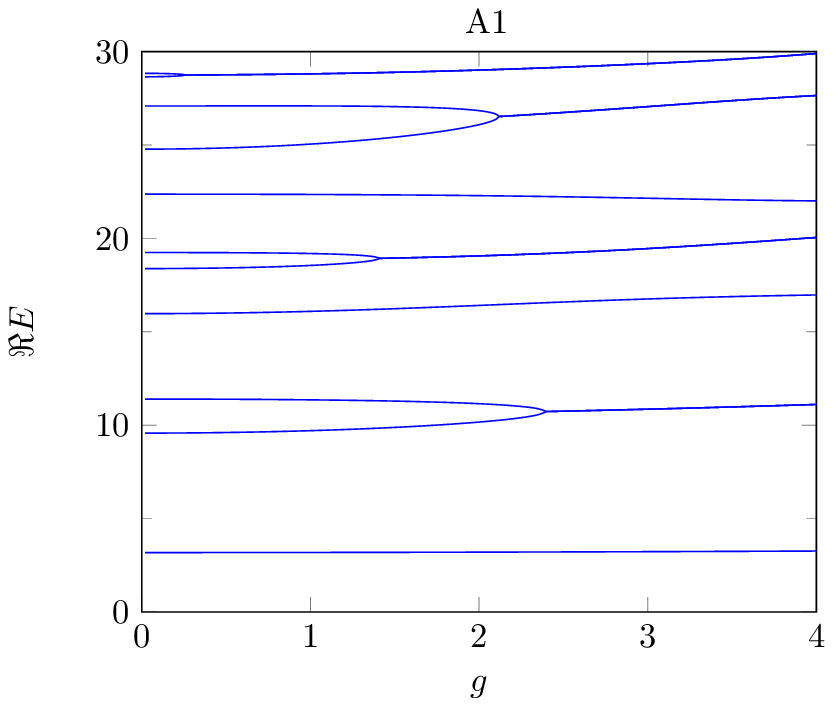} %
\includegraphics[width=6cm]{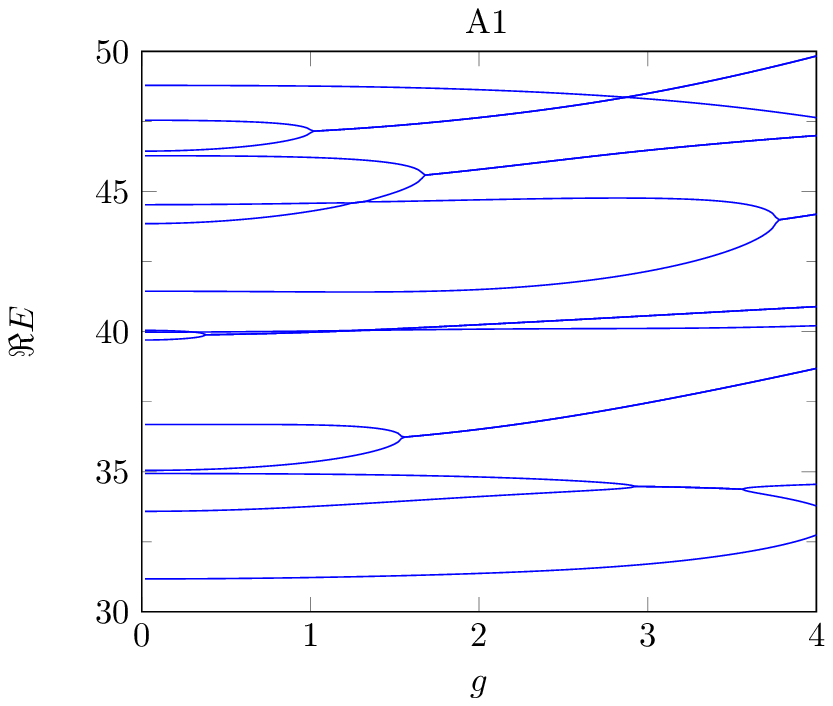} %
\includegraphics[width=6cm]{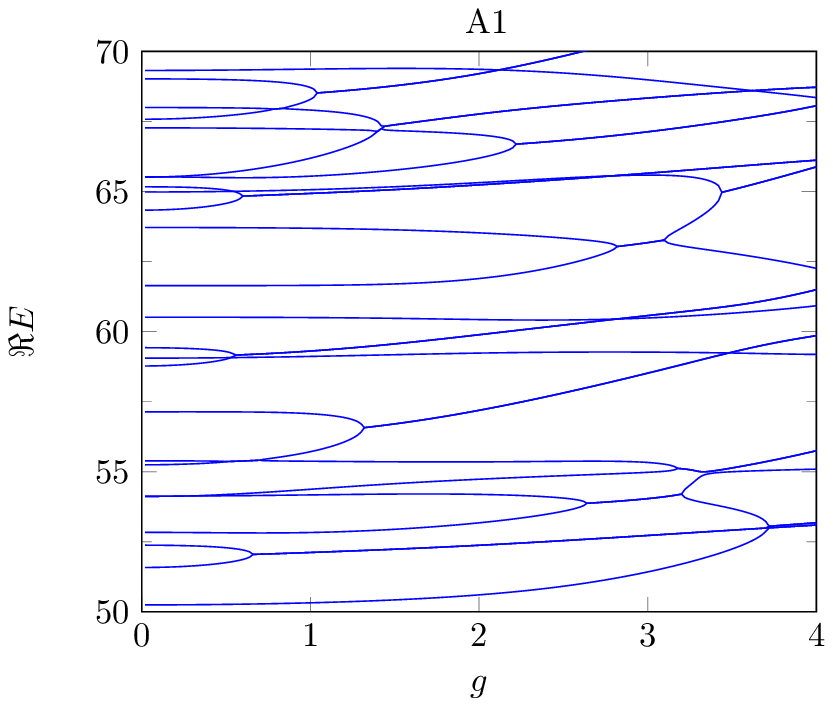}
\par
\end{center}
\caption{Real parts of the eigenvalues of symmetry $A_1$ of the Hamiltonian
operator $H=p_x^2+p_y^2+p_z^2+x^4+y^4+z^4+igxyz$}
\label{fig:Q_A1}
\end{figure}

\begin{figure}[]
\begin{center}
\includegraphics[width=6cm]{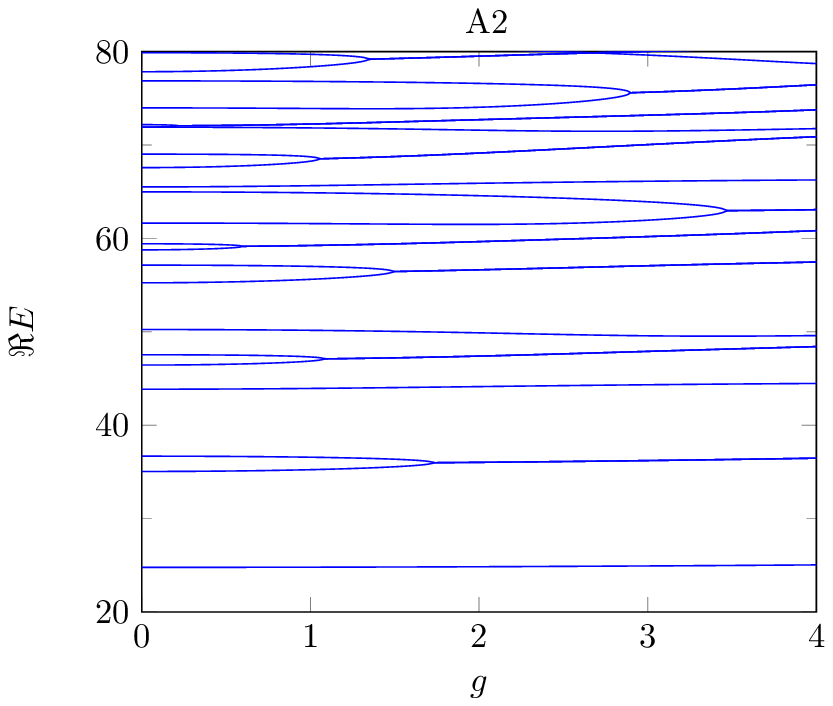} %
\includegraphics[width=6cm]{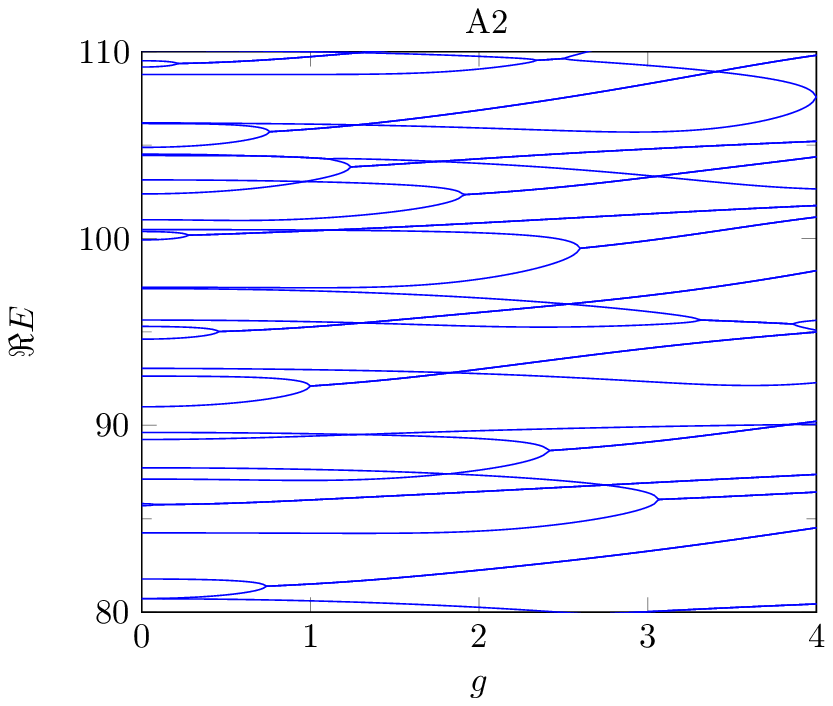}
\par
\end{center}
\caption{Real parts of the eigenvalues of symmetry $A_2$ of the Hamiltonian
operator $H=p_x^2+p_y^2+p_z^2+x^4+y^4+z^4+igxyz$}
\label{fig:Q_A2}
\end{figure}

\begin{figure}[]
\begin{center}
\includegraphics[width=6cm]{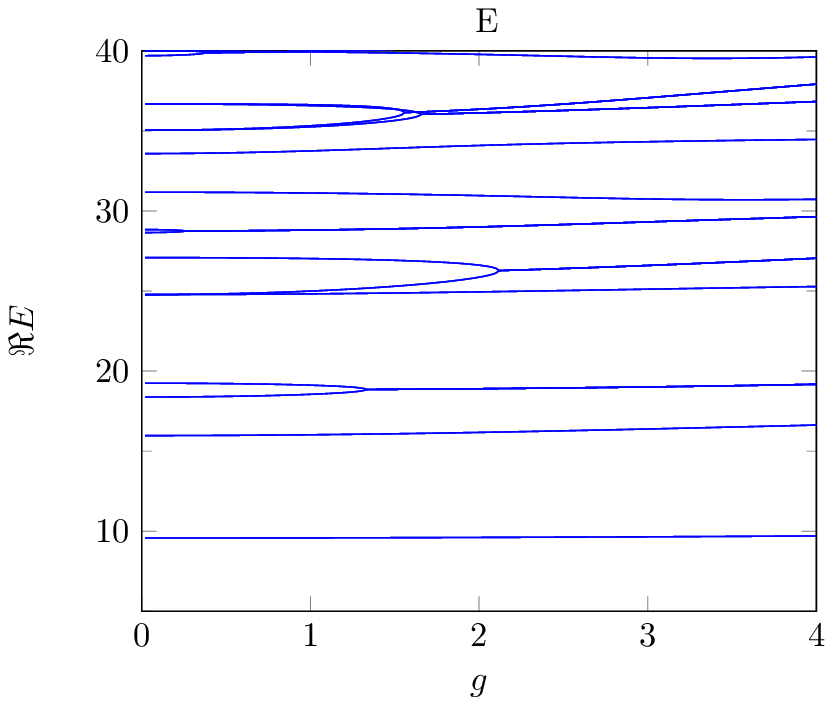} %
\includegraphics[width=6cm]{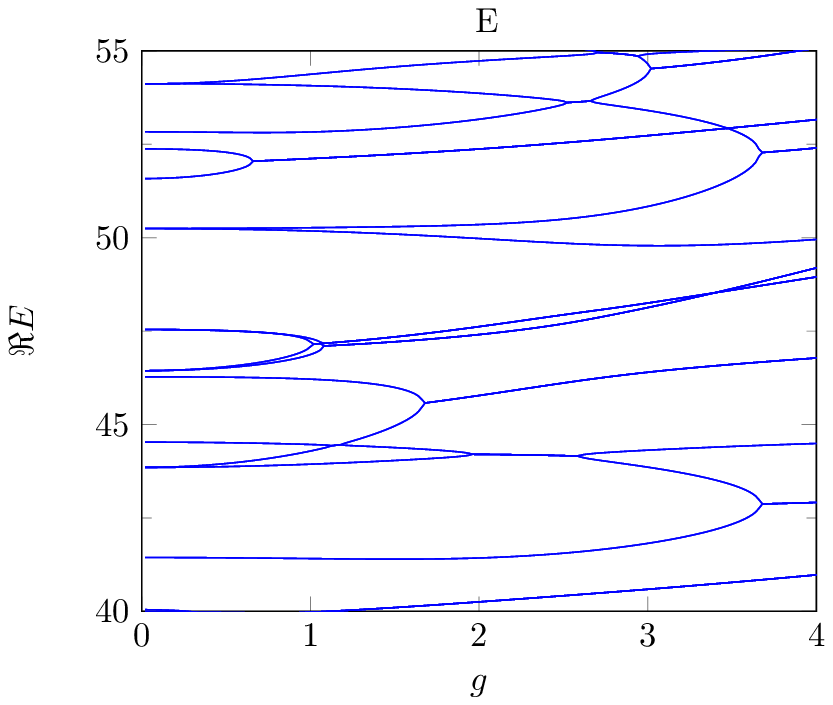} %
\includegraphics[width=6cm]{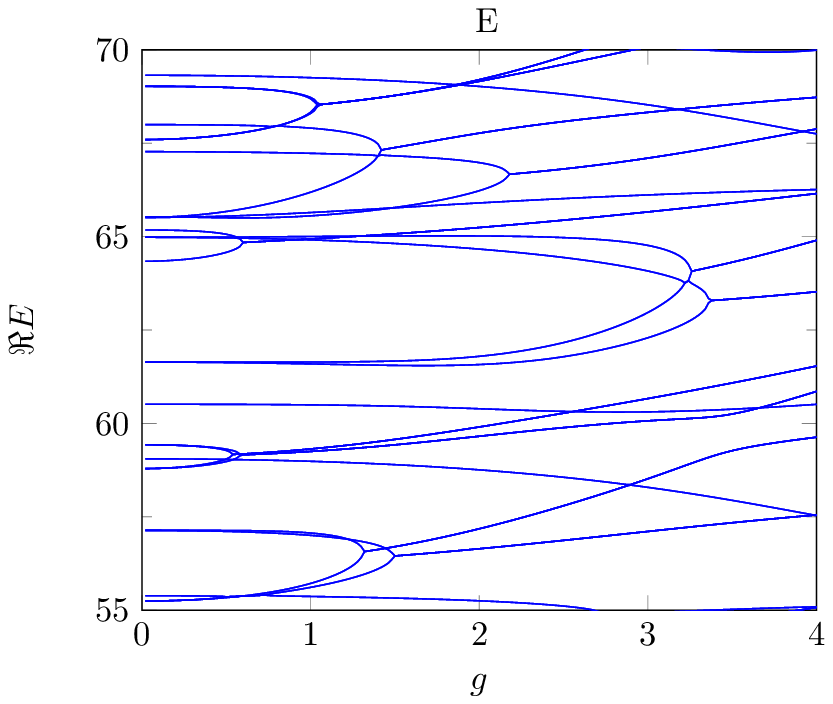}
\par
\end{center}
\caption{Real parts of the eigenvalues of symmetry $E$ of the Hamiltonian
operator $H=p_x^2+p_y^2+p_z^2+x^4+y^4+z^4+igxyz$}
\label{fig:Q_E}
\end{figure}

\begin{figure}[]
\begin{center}
\includegraphics[width=6cm]{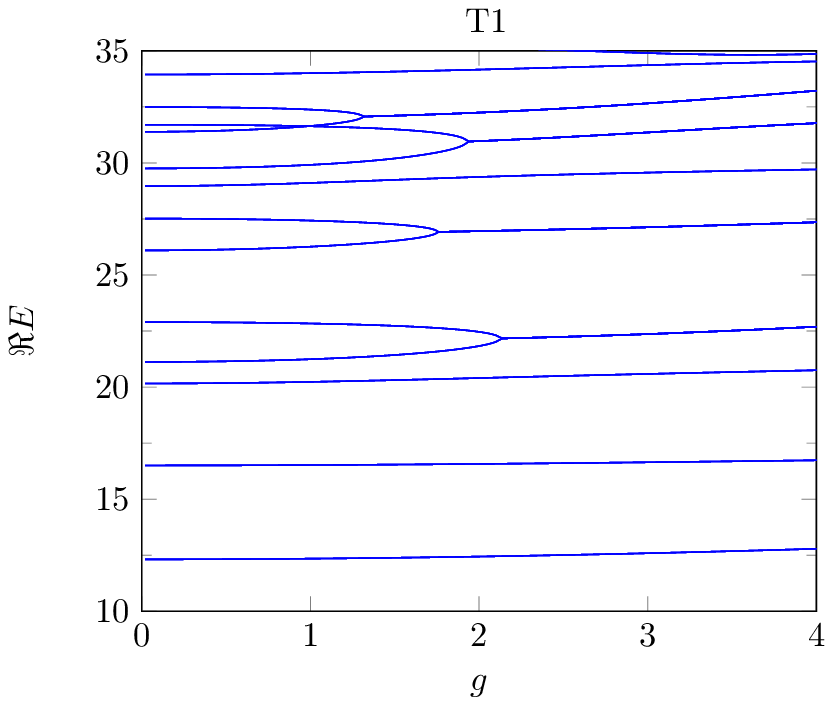} %
\includegraphics[width=6cm]{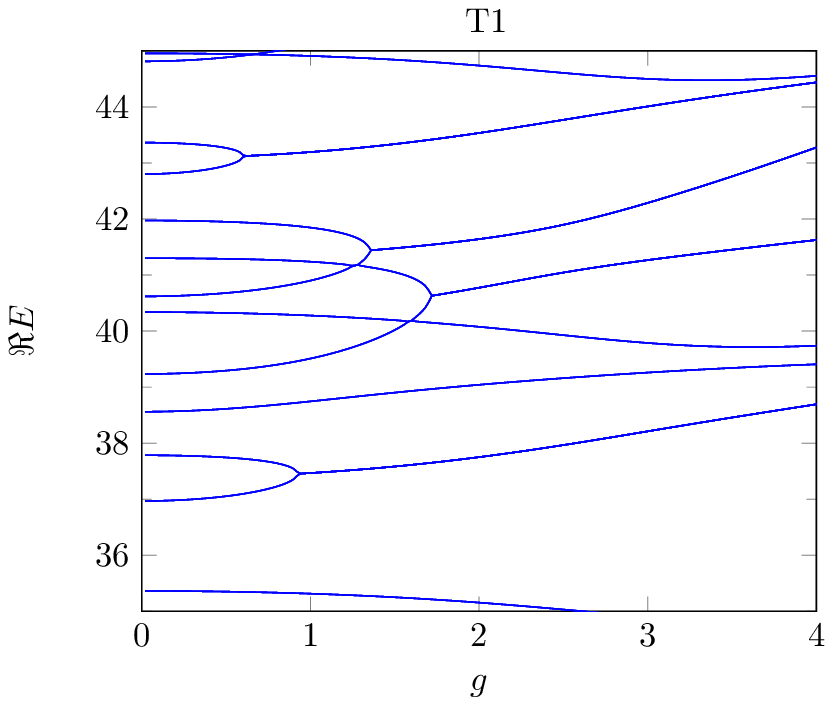} %
\includegraphics[width=6cm]{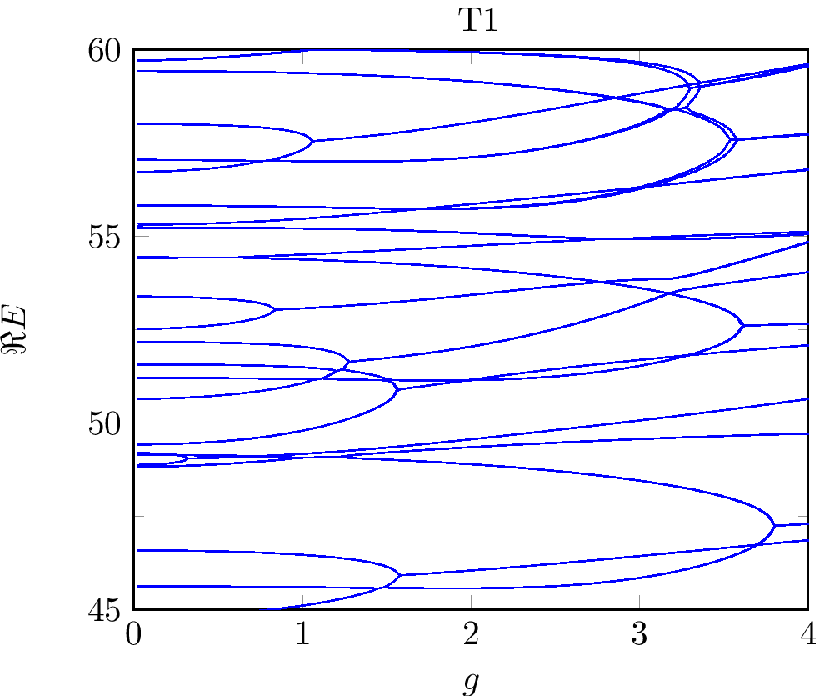}
\par
\end{center}
\caption{Real parts of the eigenvalues of symmetry $T_1$ of the Hamiltonian
operator $H=p_x^2+p_y^2+p_z^2+x^4+y^4+z^4+igxyz$}
\label{fig:Q_T1}
\end{figure}

\begin{figure}[]
\begin{center}
\includegraphics[width=6cm]{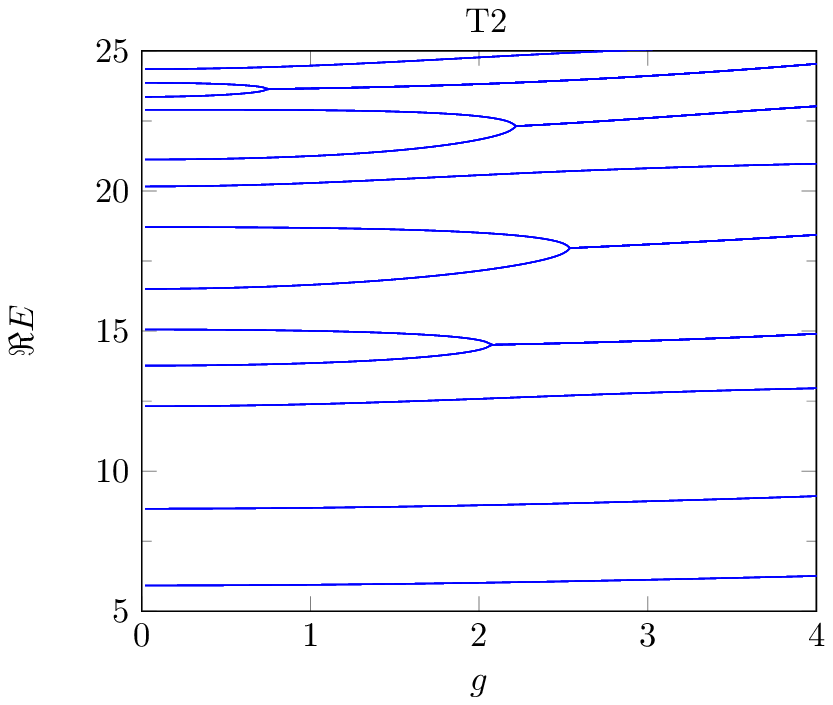} %
\includegraphics[width=6cm]{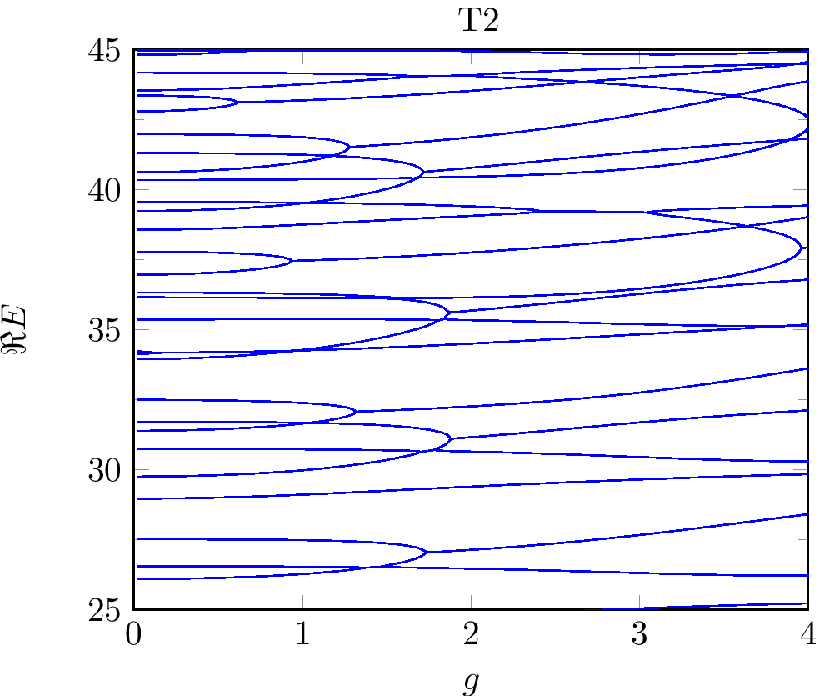}
\par
\end{center}
\caption{Real parts of the eigenvalues of symmetry $T_2$ of the Hamiltonian
operator $H=p_x^2+p_y^2+p_z^2+x^4+y^4+z^4+igxyz$}
\label{fig:Q_T2}
\end{figure}

\begin{figure}[]
\begin{center}
\includegraphics[width=6cm]{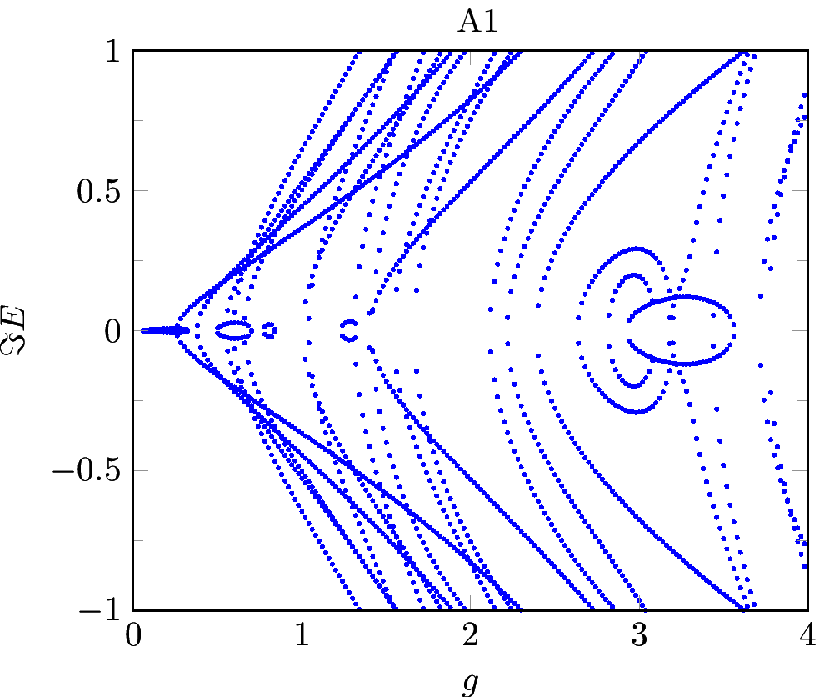} %
\includegraphics[width=6cm]{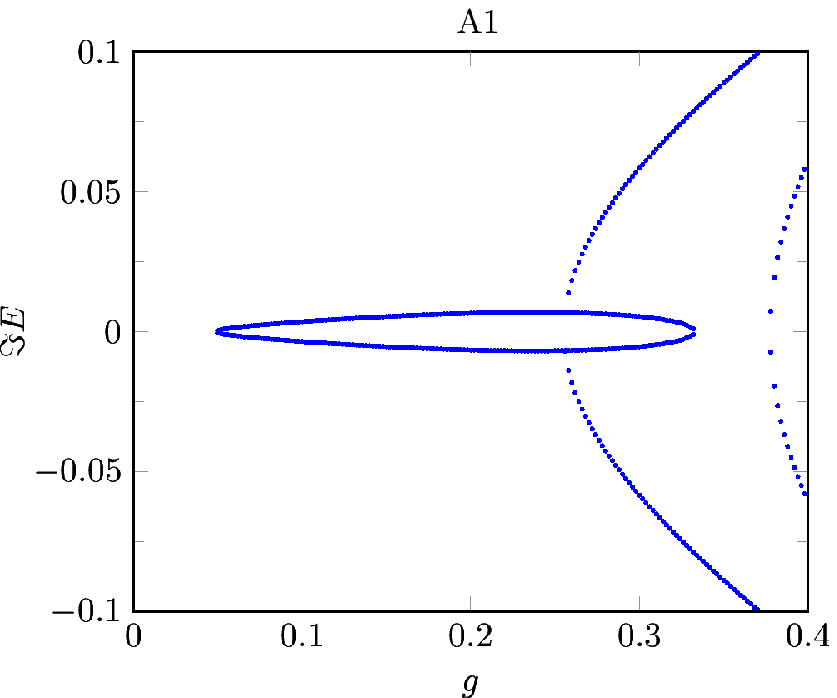} %
\includegraphics[width=6cm]{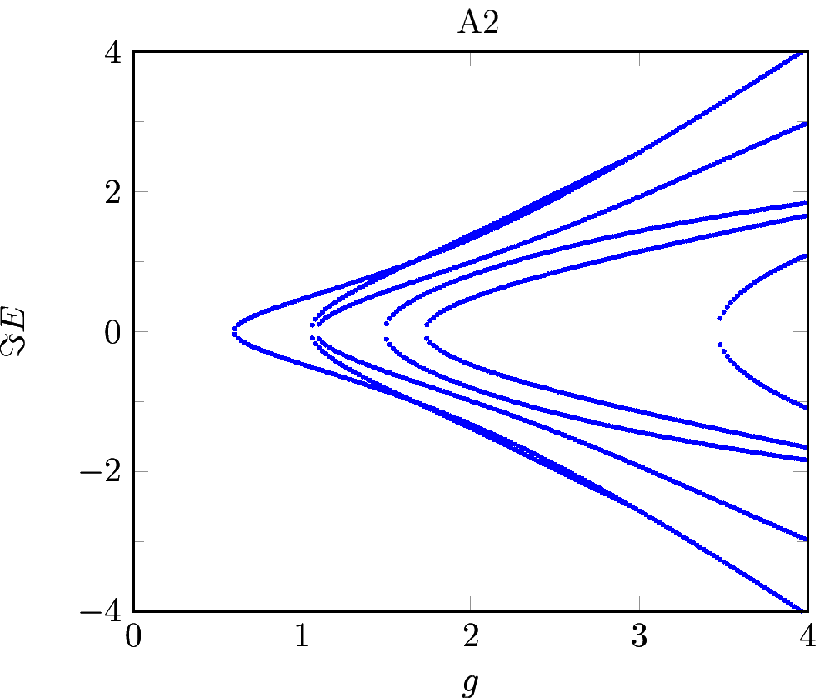} %
\includegraphics[width=6cm]{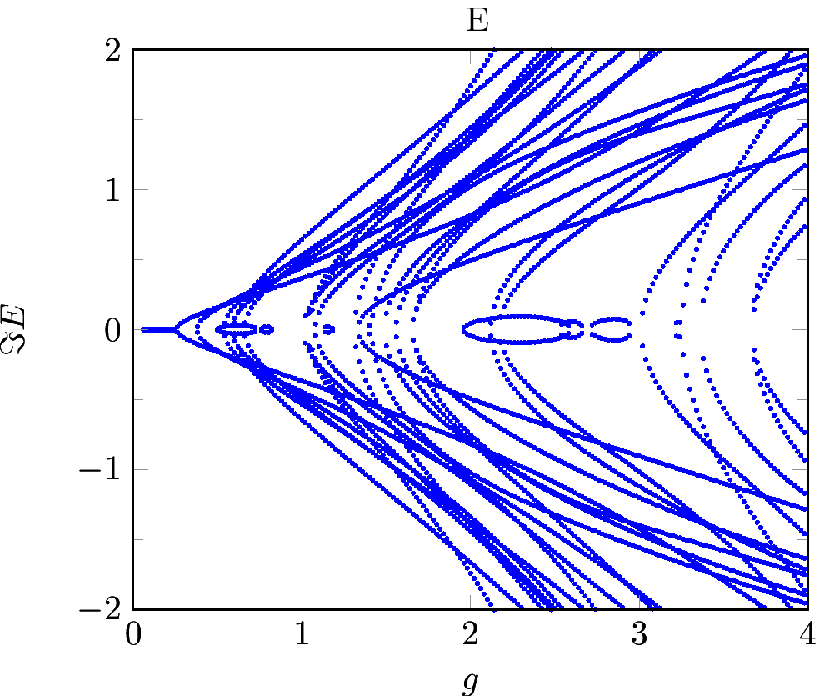} %
\includegraphics[width=6cm]{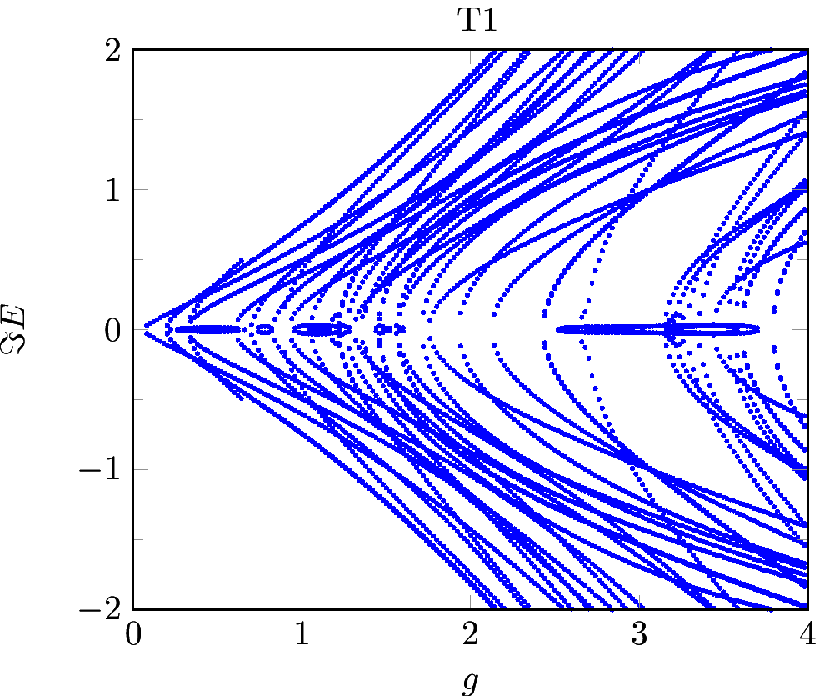} %
\includegraphics[width=6cm]{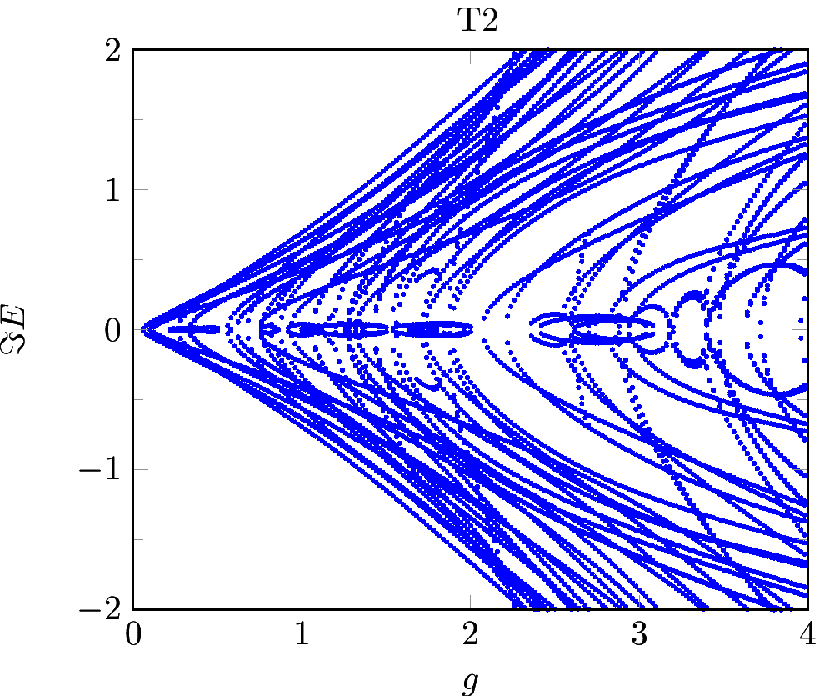}
\par
\end{center}
\caption{Imaginary parts of the eigenvalues of the Hamiltonian operator $%
H=p_x^2+p_y^2+p_z^2+x^4+y^4+z^4+igxyz$}
\label{fig:Im_Q}
\end{figure}

Clearly, $A=\hat{\imath}K$ leaves $H$ invariant. Since
$\hat{\imath}$ forms a class by itself as shown by the character
table~\ref{tab:oh} then this antiunitary symmetry is expected to
be unbroken for sufficiently small values of $g$ according to the
discussion in section~\ref {sec:diagonalization}. This conclusion
is confirmed by the figures \ref {fig:Q_A1}-\ref{fig:Im_Q} where
we see that there are real eigenvalues for sufficiently small
values of $g$ for the five irreps. However, the values of $g_{c}$
approach the origin as the eigenvalues increase in such a way that
it is difficult to estimate whether there is a high-energy phase
transition. It is also worth noting that the pattern of $\Im E$ vs
$g$ is not the same for all the irreps. The most striking
difference occurs between the irreps $A_{1}$ and $A_{2}$.

\section{Conclusions}

\label{sec:conclusions}

In this paper we have studied a few examples of non-Hermitian Hamiltonian
operators of the form (\ref{eq:H_gen}) with a space-time symmetry given by
an antiunitary operator $A=U_{a}K$. The space transformation $U_{a}$
satisfies $U_{a}H_{0}U_{a}^{-1}=H_{0}$ and $U_{a}H^{\prime
}U_{a}^{-1}=-H^{\prime }$. Under such conditions our conjecture is that one
expects real eigenvalues for sufficiently small values of $|g|$ when $U_{a}$
forms a class by itself in the point group $G_{0}$ that describes the
symmetry of $H_{0}$. This conclusion is suggested by the fact that the
perturbation corrections of first order for all the energy levels vanish.
All the known examples with real spectrum already satisfy this condition\cite
{BDMS01,NA02,N02,N05,BTZ06,KC08,W09,CIN10,BW12,HV13}. On the other hand, the
recently found space-time symmetric Hamiltonians with complex eigenvalues
for $|g|>0$\cite{FG14,FG14b,AFG14} clearly violate it. Although present
proof based on PGS and perturbation theory is not as conclusive as one may
desire, at least the examples studied so far support it.

In addition to what was said above, there remains the question whether there
is a phase transition in those cases where the eigenvalues are real for $%
0<g<g_{c}$. As $E$ increases the critical values of $g$ approach
the origin and it is quite difficult to estimate if there is a
nonzero limit.

\begin{table}[]
\caption{Character table for $T_d$ point group}
\label{tab:Td}
\begin{tabular}{l|rrrrr|l|l}
$T_d$ & $E$ & $8C_3$ & $3C_2$ & $6S_4$ & $6\sigma_d$ &  &  \\ \hline
$A_1$ & 1 & 1 & 1 & 1 & 1 &  & $x^2+y^2+z^2$ \\
$A_2$ & 1 & 1 & 1 & -1 & -1 &  &  \\
$E$ & 2 & -1 & 2 & 0 & 0 &  & $(2z^2-x^2-y^2, x^2-y^2)$ \\
$T_1$ & 3 & 0 & -1 & 1 & -1 & $(R_x, R_y, R_z)$ &  \\
$T_2$ & 3 & 0 & -1 & -1 & 1 & $(x, y, z)$ & $(xz, yz, xy)$%
\end{tabular}
\end{table}

\begin{table}[]
\caption{Character table for $O_h$ point group}
\label{tab:oh}{\tiny
\begin{tabular}{l|rrrrrrrrrr|l|l}
$O_h$ & $E$ & $8C_3$ & $6C_2$ & $6C_4$ & $3C_2(=C_4^2)$ & $i$ & $6S_4$ & $%
8S_6$ & $3\sigma_h$ & $6\sigma_d$ &  &  \\ \hline
$A_{1g}$ & 1 & 1 & 1 & 1 & 1 & 1 & 1 & 1 & 1 & 1 &  & $x^2+y^2+z^2$ \\
$A_{2g}$ & 1 & 1 & -1 & -1 & 1 & 1 & -1 & 1 & 1 & -1 &  &  \\
$E_g$ & 2 & -1 & 0 & 0 & 2 & 2 & 0 & -1 & 2 & 0 &  & $(2z^2-x^2-y^2,
x^2-y^2) $ \\
$T_{1g}$ & 3 & 0 & -1 & 1 & -1 & 3 & 1 & 0 & -1 & -1 & $(R_x, R_y, R_z)$ &
\\
$T_{2g}$ & 3 & 0 & 1 & -1 & -1 & 3 & -1 & 0 & -1 & 1 & $(xz, yz, xy)$ &  \\
$A_{1u}$ & 1 & 1 & 1 & 1 & 1 & -1 & -1 & -1 & -1 & -1 &  &  \\
$A_{2u}$ & 1 & 1 & -1 & -1 & 1 & -1 & 1 & -1 & -1 & 1 &  &  \\
$E_u$ & 2 & -1 & 0 & 0 & 2 & -2 & 0 & 1 & -2 & 0 &  &  \\
$T_{1u}$ & 3 & 0 & -1 & 1 & -1 & -3 & -1 & 0 & 1 & 1 & $(x, y, z)$ &  \\
$T_{2u}$ & 3 & 0 & 1 & -1 & -1 & -3 & 1 & 0 & 1 & -1 &  &
\end{tabular}
}
\end{table}

\end{document}